\UseRawInputEncoding
\documentclass[11pt, a4paper]{article}
\usepackage{graphicx} 
\usepackage{cite}
\usepackage{graphicx}
\usepackage[graphicx]{realboxes}
\usepackage{algorithm}
\usepackage{amsmath}
\usepackage{geometry}
\usepackage{authblk}
\usepackage{setspace}
\usepackage{indentfirst}
\newcommand{\enabstractname}{\large Abstract}
\newenvironment{enabstract}{%
\par\small
\noindent\mbox{}\hfill{\bfseries\enabstractname}\hfill\mbox{}\par
\vskip 2.5ex}{\par\vskip2.5ex}

\title{The Constraining Capability of BNS Dark Sirens Observed by the LIGO Gravitational Wave Detector on the Hubble Constant}

\author[1,2]{\small Xu Chao }
\affil[1]{\small Astronomical Society of WIT, Wuhan Institute of Technology, Wuhan 430205,China}
\affil[2]{\small School of Chemical Engineering and Pharmacy, Wuhan Institute of Technology, Wuhan 430205,China}
\date{}

\begin{document}

\maketitle

\begin{enabstract}
    The Hubble Constant observed at high redshift and low redshift are inconsistent, representing one of the urgent issues to be resolved in the field of cosmology. The discovery of gravitational waves opens a new window for addressing this problem. For instance, the GW170817 event, through the coordinated observation of electromagnetic and gravitational wave signals, allows for constraints to be imposed from a completely new perspective. However, the number of gravitational wave events where both electromagnetic and gravitational wave signals are observed simultaneously is too small, making it difficult to enhance the precision through statistical methods. In this paper, we use dark sirens as the subjects of study. Through the standard gravitational wave data simulation and the analysis process, we analyze the constraints a typical binary neutron star merger event can place on the Hubble Constant. We simulated a random event and found that it an provide an error of +0.04-0.05 for the Hubble Constant. By combining multiple events, this constraint can be improved.

\end{enabstract}

\section{Introduction} \label{sec:intro}
    In 1929, Edwin Hubble unveiled a groundbreaking discovery, establishing through the study of the distance-redshift relationship of galaxies that the universe is in a state of continuous expansion\cite{1}. He formulated the relationship between the recession velocity of galaxies and their distance as \(v = H_0r\), where \(H_0\) is known as the Hubble constant, a pivotal parameter in modern cosmology denoting the rate of cosmic expansion.

    The precise constraint of the Hubble constant can be ascertained through various methodologies, including employing the distance ladder involving Cepheid variables and Type Ia supernovae, among other observational techniques\cite{2,3,4,5}. However, a notable discrepancy has surfaced concerning the values obtained from different approaches. Current measurements of the Hubble constant fluctuate significantly between 67.4 and 76.8 $\rm{km} \cdot \rm{ s^{-1} } \cdot \rm{ Mpc^{-1} }$\cite{6}, a variation that vastly surpasses the standard uncertainty. This discrepancy, termed the ``Hubble Tension'', poses a substantial challenge to the existing cosmological paradigm, hinting at the potential necessity for new physics beyond the standard cosmological model or perhaps attributing to measurement errors\cite{7}. Numerous studies are fervently working to enhance the precision of these measurements, aspiring to delineate a more definitive conclusion in the foreseeable future\cite{8,9}.
    
    The advent of gravitational wave (GW) astronomy offers a fresh and promising avenue for study astronomy, cosmology, and physics\cite{10,11}. Unlike electromagnetic signals, gravitational waves are not obstructed by matter, potentially relaying rich information from distant astronomical entities. Moreover, gravitational wave observation raised a new method in measuring the luminosity distance of cosmic sources. It will becomes a powerful tool if the corresponding electromagnetic (EM) signal is observed. On 17 August 2017 the Advanced LIGO and Virgo detectors observed GW170817\cite{12} and its electromagnetic counter part GRB 170817A\cite{13}, they determine the Hubble constant to be $70.0^{+12.0}_{-8.0} \rm{km} \cdot \rm{ s^{-1} } \cdot \rm{ Mpc^{-1} }$\cite{14}. However, only a fraction of these compact binary sources have EM counter part, making it hard to enhance the precision through statistical methods. Fortunately, we can still extract the redshift information from binary neutron stars(BNS)\cite{15}. Such BNS gravitational wave sources without electromagnetic counterparts as ``BNS dark sirens''. To fully harness the potential of dark sirens, innovative approaches are needed to measure redshift information. In 2012, Messenger and Read proposed that the tidal deformations of neutron stars could contribute additional information to the phase evolution of gravitational waveforms, breaking the degeneracy between mass and redshift\cite{15}. This allows for the acquisition of redshift information from binary neutron star merger events exclusively through gravitational wave observations, facilitating the establishment of a relationship between luminosity distance (\(D_L\)) and redshift (\(z\)), and subsequently constraining the Hubble constant.

    This study predominantly focuses on the gravitational wave dark sirens resulting from BNS mergers, endeavoring to foresee the constraints this method can impose on the Hubble constant through predictions based on the LIGO gravitational wave detector. The paper is structured as follows: Section \ref{sec:theory} delineates the fundamental theories and computational methods concerning the binary neutron star gravitational wave dark sirens; Section \ref{sec:method} Section 3 introduces the pertinent methods and tools for gravitational wave parameter analysis and discusses the error analysis of the computed results by Bayesian methods in this section; finally, Section \ref{sec:conclusion} encapsulates the entire study. Given the scarcity of gravitational wave events, this study leverages simulated data based on the LIGO detector for analysis. To facilitate computation, we adopt a unit system where $c = G = 1$.

\section{Theory} \label{sec:theory}
    According to the theory of general relativity, the gravitational waves generated by the merger of binary compact object can be described by  two independent components, \( h_+ \) and \( h_\times \) in transverse-traceless gauge. In the frequency domain, the response of gravitational waves can be described by these two independent components\cite{16}, 
\begin{equation}
    h(f) = F_+(\theta;\phi;\psi)h_+(f) + F_\times(\theta;\phi;\psi)h_\times(f) \label{eq:1}
\end{equation}
    where \( F_+ \) and \( F_\times \) are the antenna pattern functions of the detector, (\(\theta\), \(\phi\)) represent the angles indicating the position of the gravitational wave source in the sky relative to the detector, and \( \psi \) is the polarization angle of the gravitational wave. The components \( h_+(f) \) and \( h_\times(f) \) can be obtained through post-Newtonian approximations or numerical relativity methods. A significant amount of work has been dedicated to the research of waveform construction\cite{17,18,19}. Generally, for binary black hole mergers, \( h_{+/\times}(f) \) are functions of $\Vec{\theta}$ = $ \Vec{\theta} ( m_i, a_i, 
    \theta_i, \phi_{jl}, \phi_{12}, \theta_{jn}, D_L, \psi, \phi, t_0, \alpha, \delta)$, where $m_i$ is the detector-frame mass of the $i$-th object, $a_i$ is the dimensionless spin magnitude of the $i$-th object,  $\theta_i$ is the zenith angle between the spin and orbital angular momenta for the ith object ( where i = 1, 2), $\phi_{jl}$ is the difference between total and orbital angular momentum azimuthal angles, $\phi_{12}$ is the difference between the azimuthal angles of the individual spin vector projections onto the orbital plane, $\theta_{jn}$ is the zenith angle between the total angular momentum and the line of sight, $D_L$ is the luminosity distance to the source, $\psi$ is the polarization angle of the source, $\phi$ is the binary phase at a reference frequency, $t_0$ is the GPS reference time at the geocenter (typically merger time), $\alpha$  and $\delta$ are right ascension and declination of the source.

     For the merger of binary neutron stars, an additional parameter \( \lambda \) is introduced to quantify the equation of state properties of neutron stars. During their orbital motion, the neutron stars experience tidal deformations due to tidal forces, which contribute additional phases to the gravitational waves\cite{15} (ignoring the spin of the neutron stars),
\begin{equation}
    \Psi(f)=\Psi_{\text{BBH}}(f) + \Psi_{\text{tidal}}(f)
    \label{eq:2}
\end{equation} 
    where \( \Psi_{\text{BBH}}(f) \) is the phase term in the waveform for binary black holes, 
\begin{equation}
    \Psi_{\text{tidal}}(f) = \sum_{j=1}^2 \frac{ 3\lambda_j (1+z)^5 }{128\eta M^5} \left[ -\frac{24}{\chi_j} (1 + \frac{11\eta}{\chi_j} v^5)  -\frac{5}{28\chi_j}(3179 - 919\chi_j -2286\chi_j^2 + 260\chi_j^3) v^7 \right]
    \label{eq:3}
\end{equation}

    In this work, we choose the SLy model\cite{20}, which aligns well with current observations, as the reference equation of state for neutron stars\cite{21}, and use a linear fitting function to relate the tidal deformability \( \lambda \) to the neutron star mass \( m \),
\begin{equation}
    \lambda_j = B m_j + C 
    \label{eq:4}
\end{equation} 
     Here, \( B \) and \( C \) are two tidal effect parameters; in this paper, we take \( B = -1.99 \) and \( C = 4.46 \)\cite{22}. This equation encodes redshift information into the gravitational wave waveform. In subsequent parameter inference, we can infer \( D_L \) and \( z \) simultaneously. Knowing both \( D_L \) and \( z \), we can constrain the Hubble Constant using cosmological models.

\section{Method}\label{sec:method}
    The gravitational wave signals detected by the LIGO (Laser Interferometer Gravitational-Wave Observatory) detectors are typically buried within noise\cite{23}. To extract the signals, we often employ a method called Match-Filter\cite{24}.In this work, we utilize the IMRPhenomD waveform to generate signals\cite{25,26} and incorporate the additional phase introduced in the previous section into the waveform as our gravitational wave signal $s(t)$.We generate simulated noise $n(t)$ using the power spectral density (PSD) of LIGO.The noise is then added to the simulated gravitational wave signal, resulting in the simulated data$d(t)=n(t)+s(t)$. Standard gravitational wave signal processing techniques are applied to process the simulated data, followed by Bayesian analysis to extract the parameters from the gravitational wave signal.

    For computational convenience, we have chosen a binary system of $m_1=m_2=1.4M_\odot$ for the gravitational wave event, which is also a typical neutron star mass\cite{27}.Due to the fact that neutron star mergers typically occur at the end of neutron star evolution, we assume that their spin is zero. In this work, we consider gravitational wave sources without electromagnetic counterparts, so we assume $\theta_{\rm jn}=0$. The luminosity distance $D_L$ is choosed as $40$Mpc, which is about the distance of GW170817\cite{12}. We randomly select parameters such as position, polarization, and phase to generate simulated gravitational wave signals and the redshift used for the injection is calculated using the results of Planck18 as the fiducial cosmological model.

    In the parameter inference of gravitational waves, we employ Bayesian analysis to extract the parameters from the gravitational wave signals\cite{28,29,30}:
\begin{equation}
    p(\Vec{\theta}|d) = \frac{ \mathcal{L}(\Vec{\theta}) \pi(\Vec{\theta})}{\mathcal{Z}}
    \label{eq:5}
\end{equation}
    where $\pi(\theta)$ represents the prior information, $\mathcal{L}(\theta)$represents the likelihood function, $p(\theta|d)$ is the posterior probability density, and $\mathcal{Z}$ is a normalization constant.

    In our work, for computational convenience, we fix other parameters except luminosity distance, redshift, and merger time. We will complete the full Bayesian analysis in future work.Our results are shown in Figure 1.
 From this figure, it can be observed that the data processing and parameter inference methods can provide good constraints on the luminosity distance $D_L = 38.67^{+2.85}_{-2.41} $, but they are unable to provide accurate estimates for the redshift. We obtained a relatively large error in the inferred redshift, which is $z = 0.02^{+0.03}_{-0.02}$。This is consistent with our expectations, as the sensitivity of LIGO detectors is currently not sufficient to accurately obtain redshift information using this method.
    
\begin{figure}[H]
    \centering
    \includegraphics[width=7.5cm]{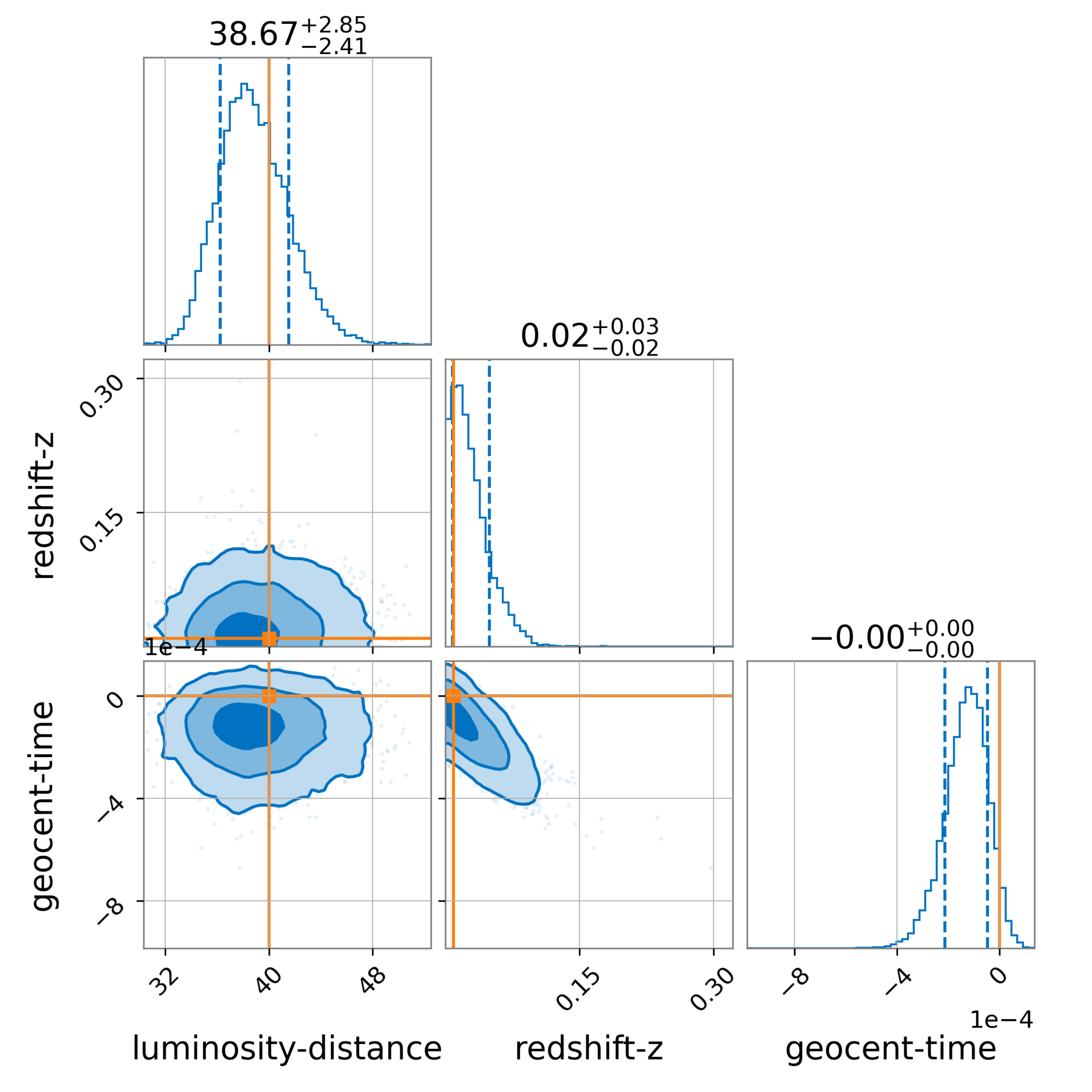}
        \caption{The results of Bayesian analysis}
        \label{fig:fast_injection_0_corner}
\end{figure}

    Once the redshift and luminosity distance are known, the Hubble constant can be calculated directly by $H_0 = v / r = zc / D_L$, where $c$ represents the speed of light, $z$ refers to the redshift of the source, and $D_L$ represents the luminosity distance of the source. The result is shown in Figure 2. In this figure, we can see the uncertainty is very small, i.e., $67.24^{+0.04}_{-0.05} \rm{km} \cdot \rm{ s^{-1} } \cdot \rm{ Mpc^{-1} }$.

\begin{figure}[H]
    \centering
    \includegraphics[width=9cm]{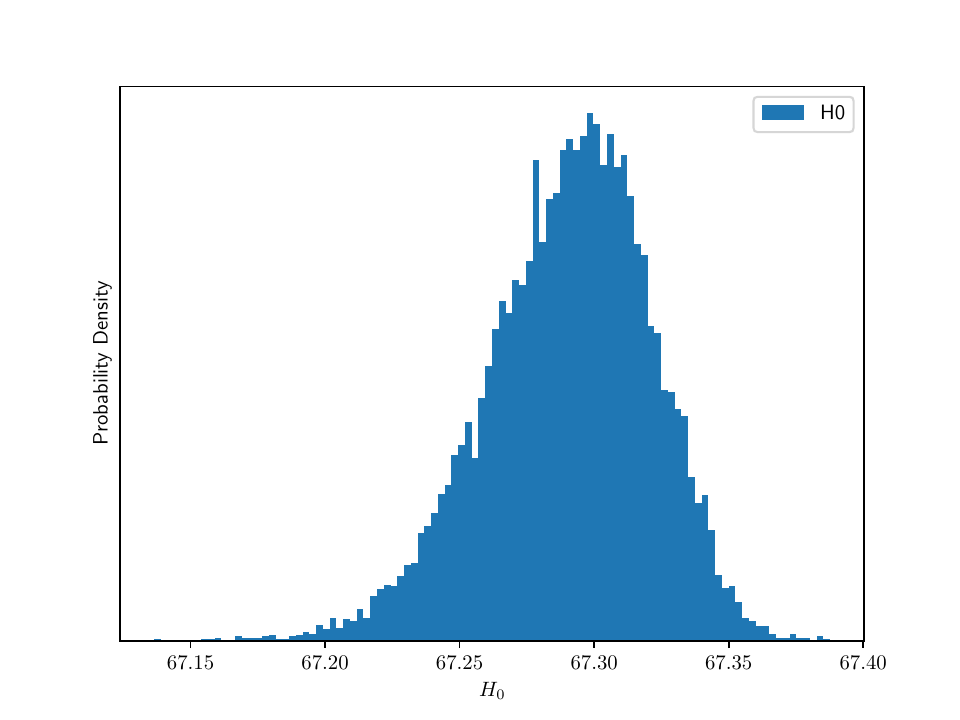}
        \caption{The uncertainty of Hubble constant}
        \label{fig:H_0}
\end{figure}

\section{Conclusion} \label{sec:conclusion}
    Gravitational waves from neutron star mergers events encode absolute luminosity distance and redshift chirp mass in their amplitude, while additional contributions in the phase can decouple the chirp mass from the redshift. This allows establishing a relationship between luminosity distance ( $D_L$ ) and redshift ($z$), thereby constraining the Hubble constant ($H_0$). In this paper, we studied the observational method based on gravitational waves from binary neutron star mergers and provided predictions for the constraining power of this method on the Hubble constant.We considered the LIGO-Hanford (H1) and LIGO-Livingston (L1) detector network, simulated gravitational wave signals from binary neutron star mergers using their instrument parameters, and used the Injection-Inference method to study the constraining power of LIGO on the Hubble Constant.

    Our results indicate that, assuming the knowledge of parameters other than luminosity distance, redshift, and merger time, the two detectors of LIGO possess strong observational properties for the Hubble Constant. However, the impact of other parameters on the Hubble Constant warrants further exploration in subsequent articles.
    
    In the future, with the construction of third-generation gravitational wave detectors, we will be able to detect a larger number of gravitational wave signals from binary neutron star mergers.The BNS dark sirens will still play an important role. We expect that in the future, BNS dark sirens will find widespread applications and provide valuable constraints on the Hubble Constant. These independent observations, separate from electromagnetic radiation, will complement existing observations and contribute to a more comprehensive understanding of the universe.

\bibliographystyle{unsrt}
\bibliography{ref}
\end{document}